\documentclass[fleqn,twoside]{article}
\usepackage{espcrc2}

\usepackage{graphicx}
\usepackage[figuresright]{rotating}
\usepackage{hyperref}
\usepackage{ifthen}
\usepackage{bm}

\newcommand{\AmS}{{\protect\the\textfont2
  A\kern-.1667em\lower.5ex\hbox{M}\kern-.125emS}}
 
\title{Acceleration from Modified Gravity: Lessons from Worked Examples}

\author{Wayne Hu\address[KICP]{Kavli Institute for Cosmological Physics, Department of Astronomy \& Astrophysics, Enrico Fermi Institute, University of Chicago, Chicago, IL 60637}%
        \thanks{This work was supported by the KICP through grants NSF PHY-0114422 and
 NSF PHY-0551142,  U.S.~Dept.\ of Energy contract DE-FG02-90ER-40560 and
 the David and Lucile Packard Foundation.
 I thank W. Fang, M. Lima, L. Lombriser,  H. Oyaizu, I. Sawicki, F. Schmidt, YS. Song, A. Upadhye for fruitful collaborations;
  L. Hui, J. Khoury, K. Koyama, B. Wald for illuminating discussions; GGI and the organizers of the New Horizons workshop for their
  hospitality.}}
       
\begin{document}

\begin{abstract}
I examine how two  specific examples of modified gravity explanations of cosmic acceleration
 help us understand some general
problems confronting cosmological tests of gravity: how do we distinguish
 modified gravity from dark energy if they can be made formally equivalent?
 how do
 we parameterize deviations according to physical principles with
 sufficient generality,
yet focus cosmological tests into areas that complement our existing knowledge of
 gravity? how do we treat the dynamics of modifications which necessarily involve non-linearities that preclude
 superposition of forces?  
The modified action $f(R)$ and DGP braneworld models provide insight on these question as fully-worked examples whose
 expansion history, linear perturbation theory, and most recently,  non-linear $N$-body and force-modification field
 dynamics of cosmological simulations are available for study.
\vspace{1pc}
\end{abstract}

\maketitle

\section{INTRODUCTION}

The next generation of surveys aimed at characterizing the acceleration of the
expansion will also provide precision tests of gravity in a cosmological
context.   In particular, they will address the possibility that the acceleration itself
is driven not by an unknown dark energy component but rather by a modification to
gravity on the largest scales (e.g.~\cite{Albrecht:2006um,Albrecht:2009ct,Frieman:2008sn,Linder:2008pp}).

In the absence of a compelling theory for modified gravity, how do we best
quantify and focus these tests?   Is there an analogue of the parameterized
post-Newtonian description of local gravity for cosmology? Can it act
as a meeting point between observations and theory? 

 At the very outset
there are some basic questions that confront this endeavor.  In particular,
three issues, loosely based on the challenges compiled in \cite{Linder:2008pp},
appear as obstacles:
\begin{enumerate}

\item{} Dark energy equivalence.  Formally, any metric theory of modified gravity can
be recast into an equivalent dark energy theory under ordinary gravity.  Is a cosmological test of
gravity itself even possible?

\item{} Anything goes.   Without the framework of general relativity we are left with
arbitrary functions of space and time to relate the matter and the metric.  What guidance
do we have to parameterize them?

\item{} Superposition.   Order unity cosmological deviations and tiny local deviations 
imply that the modified Poisson equation is non-linear in some way.  How do we
treat non-linear dynamics without 
a superposition principle for gravitational forces?

 \end{enumerate}
Given such basic issues, it is useful to have concrete examples, however contrived, ad hoc,
or problematic on a quantum level.  They serve as case studies for how a modified gravity theory might work.  

 Two such examples are the modified action model   where the Einstein-Hilbert action is supplemented by a non-linear function of the Ricci scalar
 $f(R)$ \cite{Caretal03,NojOdi03,Capozziello:2003tk}
 and the 5-dimensional braneworld model of Dvali, Gabadadze, \& Porrati
 (DGP) \cite{DvaGabPor00}.   Both cases are now fully worked in 
 that their expansion history, linear perturbation theory, and most recently,  non-linear $N$-body
 dynamics of cosmological simulations have been studied.  

In this piece, I examine how these worked examples provide insight on the three conundrums  confronting cosmological tests of gravity for cosmic acceleration.

\section{DARK ENERGY EQUIVALENCE}
\label{sec:equivalence}

Given the formal equivalence which maps a modified gravity theory onto a 
dark energy theory under ordinary gravity, can cosmological observables ever test gravity itself? 

Let's examine the formal equivalence and then place it in the context of the two worked
examples.  
A metric theory of gravity where
energy-momentum is covariantly (locally) conserved ($\nabla^\mu T_{\mu\nu}=0$) can always be rewritten
as a dark energy theory.   To see this, consider that a generally covariant field equation
of the form (e.g.~\cite{Kunz:2006ca,HuSaw07b})
\begin{equation}
S_{\mu\nu}(g_{\mu\nu}) = 8\pi G T_{\mu\nu}
\end{equation}
can be re-expressed as ordinary gravity $G_{\mu\nu}=8\pi G T_{\mu\nu}$ with
an effective dark energy stress energy
\begin{equation}
T_{\mu\nu}^{\rm DE} \equiv {1 \over 8\pi G}( G_{\mu\nu}-S_{\mu\nu}) \,.
\end{equation}
Moreover this effective stress-energy is covariantly conserved 
$\nabla^\mu T_{\mu\nu}^{\rm DE}=0$ by
virtue of the Bianchi identity.    Also, conservation of the matter means that
the only influence of the modification to gravity or effective dark energy comes through the
metric: once the gravitational potentials are determined, matter falls in them
as usual.

 While this formal equivalence holds, it does not mean that
 the equivalent dark energy description obeys reasonable microphysics.
 Microphysics determine the equations of state that relate the components of the stress energy tensor: the pressure, energy density, momentum density and anisotropic stress.    
 The effective dark energy description of a modified gravity model will
 look contrived as its microphysics must
mimic an explicit dependence on the metric unlike in 
 physical dark energy models like scalar fields \cite{HuSaw07b}.  
 
Moreover, even though the matter and effective dark energy are separately conserved,
the relationship between the effective microphysics and the metric imply an implicit coupling
between the  dark energy and matter -- i.e.~the effective dark energy mediates a fifth forces that universally couples to the matter.
The $f(R)$ and DGP models are both exhibit this phenomenology.  They contain
an extra scalar degree of freedom that can either be represented as dark energy
anisotropic stress, proportional to the metric, or the propagating mediator of a fifth force.
As such, they are phenomenologically quite distinct from typical dark energy models.

In summary, cosmological tests of gravity, just like solar system measurements, require a notion of what
composes the stress-energy tensor(s) in the system.  For cosmological tests, 
we assume candidates for the dark matter and dark energy with microphysics that,
along with conservation of energy-momentum, closes their equations of motion
\cite{Hu98}.   What most cosmologists implicitly mean by a ``test of gravity" is:
 a test of gravity in the context of cold dark matter and dark energy whose density is relatively spatially smooth 
 on 
scales well below the current horizon.\footnote{Modified gravity explanations of the dark matter fall into a different category and are not the subject of this exploration.}    In this context, general relativity makes
specific and falsifiable predictions for cosmological observables.  Falsification however
does not necessarily imply that general relativity is incorrect.

\section{ANYTHING GOES}

Without the framework of general relativity how do we parameterize and test gravity cosmologically?

\smallskip
{\it Consistency Approaches:}  If anything goes, then shouldn't one start with parameters
that define the simplest self-consistency tests
of general relativity (plus dark energy that is smooth relative to cold dark matter)?

Candidates for self-consistency tests include an approximate expression for
the growth rate of {\it linear} density perturbations $\Delta_m$ given the matter 
contribution to the expansion rate $\Omega_m(a)$ (e.g.~\cite{Linder:2007hg,HutLin07})
\begin{equation}
{d \ln \Delta_m  \over d\ln a} \approx \Omega_m(a)^\gamma,
\end{equation}
where $\gamma \approx 0.55$.  Other approaches include 
separate 
dark energy equation of states that govern growth and distance measures
 (e.g.~\cite{IshUpaSpe06,WanHuiMayHai07}) or 
model independent reconstructions of growth from distance measures that can be
compared with growth measurements
(e.g.~\cite{Mortonson:2008qy,Alam:2008at}).  The latter have the advantage that consistency
checks can be made exact in linear theory whereas the former two parametric schemes are easier to
apply. 

One
problem with these approaches is that deviations in these parameters away from
consistency do not represent
physically possible modified gravity models.
As minimal approaches, they also do not target
areas where measurements might best complement what we already know about
gravity.   For example,
consistency statistics can be dominated by information from  the non-linear regime
where, as we shall see, ordinary gravity is almost guaranteed.  Conversely, an incomplete
account of non-linear baryonic
physics can give false consistency violations \cite{Hearin:2009hz}.  

Likewise, characteristic signatures of  modified gravity aren't
very well exposed by these consistency statistics.  Neither the $f(R)$ nor DGP models
have  a scale independent linear growth rate, though deviations for the latter only
appear near the horizon scale \cite{SonHuSaw06,SawSonHu06,cardoso:08}.  Moreover modified gravity theories
such as these
change the relationship between the gravitational potential, spatial curvature
and the density perturbations.

 It is tempting to simply supplement consistency parameters
like $\gamma$, generalized perhaps to be a function of scale, with a variation in the Newton constant $G$ that also depends on scale
(e.g.~\cite{Amendola:2007rr}).  However these types of prescriptions do not 
establish a framework where physical principles like conservation of energy-momentum
strictly apply nor 
do they provide a viable construction beyond linear perturbations. 

Despite these flaws, simple consistency tests are valuable in that they are easy to
apply and forecast. Their violation will point to new physics or astrophysics that can then be
more fully developed.  They are not however the only reasonable metric to apply 
for judging tests of gravity.

\smallskip
 {\it Linear Parameterization:}
Fortunately, the problem of formal equivalence to dark energy in \S \ref{sec:equivalence}
and knowledge that
gravity must only have small deviations from general relativity locally (see \S \ref{sec:superposition})
 is a virtue for developing a parameterized description of modified gravity.
 
The framework of general
relativity and aspects of the parameterized post Newtonian description 
for deviations carry over to modified gravity.
Coupled with specific
worked
examples of the types of deviations expected in modified gravity models, this framework
 brings form to a cosmological parameterization 
of gravity.  

As discussed in \S \ref{sec:equivalence}, if we can parameterize the degrees of
freedom associated with the microphysics of the effective dark energy then we
have a description that satisfies physical principles: gravity as a metric theory and
and conservation of energy-momentum.

Modified gravity theories, including the two worked examples,
typically have an extra scalar degree of freedom associated with the physics that
drives the acceleration.  Just as in the parameterized post-Newtonian (PPN) description,
this scalar characterizes the relationship between the Newtonian potential $\Psi$ and
the spatial curvature $\Phi$ but now in the context of
 small deviations from a statistically homogeneous
and isotropic Friedmann-Robertson-Walker
background 
\begin{eqnarray}
g &=& {\Phi + \Psi \over \Phi-\Psi}\,,\\
ds^2 &=& -(1+2\Psi)dt^2 + a^2 (1+2\Phi)dx^2\,.
\end{eqnarray}
Note that $g=0$ for a universe composed of cold dark matter and smooth dark energy.   These deviations can be measured
by comparing gravitational lensing and gravitational redshifts, which are sensitive
to $\Phi-\Psi$, with the dynamics of non-relativistic matter, which is sensitive to $\Psi$. 

In terms of the effective dark energy, the metric deviation $g$ is related to anisotropic
stress (e.g. \cite{Uzan:2006mf,CalCooMel07}).  On superhorizon scales, this parameter
$g(a,k\rightarrow0)$
alone is sufficient to describe the evolution of the two potentials $\Phi$, $\Psi$ given
a known, separately parameterized, expansion history
\cite{HuEis99}.  Completeness follows from the conservation of the comoving
curvature for adiabatic perturbations
\begin{equation}
\zeta = {\rm const.}\quad (k \ll aH)
\label{eqn:zeta}
\end{equation}
independently of the microphysics of dark energy.  This independence can
also be directly shown without the aid of the dark energy equivalence 
\cite{Ber06}.   Fundamentally, it only requires that conservation
of energy-momentum still apply.

This fact is also quite useful for describing the two worked 
modified gravity examples.
In the DGP scenario, the complicated superhorizon 
physics of metric fluctuations propagating into the extra dimension through the
master equation for the bulk  is simply 
encapsulated by this one function on the brane \cite{Sawicki:2006jj}.   The inclusion of this horizon 
scale effect leads to an excess in the large angle CMB anisotropy \cite{SonSawHu06}
which brings the self-accelerating solution into $5\sigma$
tension with the joint CMB and supernovae data \cite{Fang:2008kc}.   Likewise, the
complicated 4th order dynamics of $f(R)$ theories is also neatly encapsulated in $g$
on large scales \cite{SonHuSaw06}.

On smaller scales, conservation of energy-momentum and $g$ is not sufficient for 
a complete description.  Again dark energy equivalence is useful for understanding
this fact.   There it is necessary to close the equations of motion by supplementing
conservation with an equation of state that relates the pressure fluctuation to
the  density, momentum and anisotropic stress fluctuations.  For a smooth dark
energy component, this is achieved by setting a Jeans scale or a sound speed for
pressure and anisotropic stress fluctuations for the
dark energy \cite{Hu98}.

One possibility that has been explored \cite{CalCooMel07,Daniel:2008et} is setting the
anisotropic stress and pressure of the effective dark energy to exactly cancel
so as to leave no source to the momentum density in a particular gauge choice,
synchronous gauge.   This amounts to assuming that $\zeta$ is a constant
on all scales if the remaining matter is cold dark matter and corresponds to 
a case where the effective dark energy has non-negligible density contributions 
on small scales.\footnote{The same issue applies to synchronous gauge techniques
that simply shut off dark energy perturbations on small scales as an approximation 
to smooth dark energy: in a synchronous gauge choice where the dark matter defines
the frame, the momentum density of the dark energy is never negligible.}
However it is too restrictive and
 does not lead to a small scale
description that has the features of the worked examples.  

In both the DGP
and viable $f(R)$ models, the Newtonian regime is well approximated by an 
unmodified Poisson equation for the lensing potential
\begin{equation}
\nabla^2 \left( {\Phi-\Psi \over 2} \right) = -4\pi G a^2 \bar\rho_m \Delta_m\,,
\label{eqn:poisson}
\end{equation}
where spatial derivatives are comoving throughout. 
Along with $g$ and energy-momentum conservation for the matter, the Poisson
equation for the lensing potential completes
the system.  More generally $G$ can be made time dependent without breaking
the structure of these equations \cite{HuSaw07b}.  On
the other hand, parameterizations that just involve $g$ and the Poisson equation
\cite{Jain:2007yk,Amendola:2007rr,AmiWagBla07} fail to automatically enforce energy-momentum conservation or the Bianchi
identity on horizon
scales and above.  For the $f(R)$ model $g\rightarrow -1/3$ below the Compton wavelength
scale in the background and for the DGP model $g \rightarrow -1/3\beta(a)$ where 
$\beta(a) = 1-2 Hr_c(1+ \frac{1}{3}d\ln H/d\ln a)$ with $r_c$ as the crossover scale.

A parameterized post-Friedmann (PPF) description that is flexible enough to incorporate both the large
and small scale behavior of linear fluctuations in the two worked examples for
modified gravity was introduced in \cite{HuSaw07b}.   The gist of this parameterization
is to include an interpolation between the large scale  Friedmann behavior
of equation (\ref{eqn:zeta})
and the Newtonian behavior of equation (\ref{eqn:poisson}) while strictly conserving energy-momentum
on all scales.   This amounts to  highly non-trivial closure relations or equations of
state for the equivalent effective dark energy.   Such complexity addresses the question of
distinguishability raised in \S \ref{sec:equivalence}.

These relations are readily generalized to the early universe where
radiation fields from the photons and neutrinos are important and provide their own
anisotropic stress \cite{Hu08}.  A public code to calculate linear fluctuations including the
CMB under
this generalized PPF description was
released in \cite{Fang:2008sn}.\footnote{\url{camb.info/ppf}}  This method has also
been adapted to handle smooth dark energy, braneworld gravity with brane tension
\cite{Lombriser:2009xg}, and explorations of degravitation ideas \cite{Afshordi:2008rd}.

\smallskip
{\it Non-linear Parameterization:}
There is one final, perhaps most critical, piece to a PPF description.  Given that there
exists very strong constraints on analogous PPN parameters locally, e.g.~in solar system
tests, there must be some means by which finite modifications at cosmological scales
are reduced to acceptable levels locally.    

Here is where having the worked
examples prove the most valuable.   In the DGP and $f(R)$ scenarios, ordinary gravity
is restored by having non-linear interactions of the force-modification field
in the presence of non-linear matter fluctuations. 
It is {\it not} sufficient to simply have $g(a,k)$
run with wavenumber since the restoration works in real space and is dependent
on the local environment.  The PPF approach in \cite{HuSaw07b} for 2-point statistics
takes the rms density field as an interpolation parameter between linearized modified
gravity and non-linear ordinary gravity inside collapsed dark matter halos.  This approach works quite well for DGP where
the non-linearity is linked closely with density \cite{Koyama:2009me}.
The original PPF approach requires generalization for other non-linear mechanisms which 
interpolate between the linear and deeply non-linear regime in different ways and for
statistics beyond 2-point functions

While the non-linear PPF description is still incomplete, it is crucial that a parameterized scheme incorporate at least one degree of freedom 
which allows gravity  to return to normal locally.  Without this feature one
might infer overly tight limits on modifications to gravity in the linear regime from non-linear statistics such as the cluster abundance or cosmic shear (cf. \cite{Rapetti:2008rm}).  This caveat applies to consistency parameters
like $\gamma$ as well.  We turn now to the guidance the non-linear dynamics of the
worked examples provide on these and related issues.

\section{SUPERPOSITION}
\label{sec:superposition}

Given that the return to  ordinary gravity locally necessitates some non-linear process, the superposition principle for forces among bodies no longer applies.  How then do we test gravity with
cosmological observables in the non-linear regime?  Does the lack of a superposition
principle, which implies mode coupling, invalidate even large-scale predictions based on linear theory? 

\smallskip
{\it Non-linear Field Dynamics:} 
The most critical and difficult piece of a viable description for modified gravity 
is how non-linear effects return ordinary gravity locally.   
The two worked examples $f(R)$ and DGP provide some guidance on how this can happen.    In both cases, on small scales and assuming
non-relativistic velocities for the matter, Eq.~\ref{eqn:poisson} for the lensing
potential continues to apply but the non-linear equivalent to $g$ comes from a 
modified Poisson equation
\begin{equation}
\nabla^2 \Psi = 4\pi G a^2 \bar\rho_m \Delta_m + {1\over 2}\nabla^2\phi \,,
\label{eqn:modpoisson}
\end{equation}
where the extra scalar $\phi = -df/dR$ in $f(R)$ theories and is associated with the
brane position in DGP. Note that the Poisson equation is still linear in the density fluctuation from the mean
$\Delta_m$ but gains an extra source from a scalar degree of freedom 
$\nabla^2\phi$.  

Given that the matter still moves in the metric as usual, non-relativistic particles still feel forces according to $\nabla \Psi$.  Nonetheless compared with ordinary gravity there is an enhanced gravitational (or ``fifth")
force from $\nabla \phi$. 
This extra source is set in linear theory to be $\phi = \Phi+\Psi = g(\Phi-\Psi)$ and so
should obey the equation
\begin{equation}
\nabla^2 \phi = g(a) a^2 (8\pi G \bar \rho_m \Delta_m),\,\, ({\rm linear}, k\gg aH)\,.
\label{eqn:fieldlinear}
\end{equation} 
To suppress the enhanced forces locally, the non-linear generalization should
have the form 
\begin{equation}
\nabla^2 \phi = g(a) a^2 \left( 8\pi G \bar \rho_m \Delta_m - N[\phi]\right)\,,
\label{eqn:fieldequation}
\end{equation}
where $g(a)$ is now taken to be the small scale limit of the {\it linear} relationship
between the metric potentials. 
Here $N[\phi]$ is some non-linear function of the field and its derivatives which goes
to
\begin{equation}
N[\phi] \rightarrow 8\pi G \bar \rho_m \Delta_m
\end{equation}
locally. 
In the $f(R)$ case, $N[\phi]=\delta R(f(\phi))$ and is a function of the local field alone.  Here, the
general relativistic expectation that $R=8\pi G \rho_m$ gives the minimum of the
effective potential for $\phi$.  If the minimum can be reached inside
an overdense region then force law deviations are shielded in the interior by
the so-called chameleon mechanism (e.g.~\cite{KhoWel04}).

In the DGP case
\begin{equation}
N[\phi] = {r_c^2 \over a^4}\left[ (\nabla^2 \phi)^2 - (\nabla_i \nabla_j \phi)^2 \right]\,.
\end{equation}
Here the non-linearity contains quadratic combinations of second derivatives of the field just like the Laplace
operator on the lhs of Eq.~(\ref{eqn:fieldequation}).  Detailed balance suggests that once
$\nabla^2\phi$ becomes large, the non-linear piece will be driven to cancel
$8\pi G \bar\rho_m\Delta_m$ with the field being a nearly algebraic function of
the overdensity.  In this case, the suppression is called the Vainshtein mechanism.

\smallskip
{\it N-body Techniques:}
The generic features of the non-linear dynamics from the worked examples are: the Poisson equation for $\Psi$ remains linear
in its sources of the density {\it fluctuation} and the Laplacian of the force-modification field
 with the matter falling in the metric as usual. Once
$\Psi$ is obtained other aspects of mesh-based $N$-body simulations remains
the same.

On the other hand,
the scalar source, like the density field, obeys a {\it non-linear} equation of motion and its
configuration in the presence of the density field must be solved numerically.  The
non-linear field equation can be solved with iterative relaxation techniques \cite{Oyaizu:2008sr}.
These have been successfully applied to the $f(R)$ \cite{Oyaizu:2008tb} and DGP sources of
non-linearity \cite{Schmidt:2009sg} in a particle-mesh N-body.
Inside a large overdensity, the scalar source responds non-linearly and suppresses
the additional contribution to gravitational forces.   Non-linearity implies that the gravitational
field is no longer the sum of all of the independent contributions from individual bodies,
instead there is a collective saturation of the sources of deviation from ordinary gravity.

This saturation makes scaling relations between linear theory predictions and non-linear
observables established for ordinary gravity dangerous to apply to modified gravity.  These
include relationships between the linear growth rate and the non-linear power spectrum and
halo mass function \cite{Oyaizu:2008tb,Schmidt:2008tn,Schmidt:2009sg}.

It is 
not even guaranteed that the average scalar source follows the linear theory
prediction on large
supposedly linear scales.   Lack of a superposition principle breaks the linear assumption
of averaging:  small scale non-linear density
fluctuations can affect the behavior of the large-scale field.   In a universe where even large
scale density fluctuations are composed from the correlations between non-linear
collapsed dark matter halos, this feature can invalidate the predictions of linear theory.
Likewise, the non-linear contributions to the scalar field around a collapsed body 
can alter the field contribution to the metric substantially.  As a consequence
 it can no longer be thought
of as moving as a test body in an external field.  We must solve self-consistently
for the internal and external field.

\smallskip
{\it Superposition and Saturation:}
There are important similarities and differences between the ways
$f(R)$ and DGP resolve the superposition and saturation issues.   In both cases,
$N$-body results show that large scale cosmological 
density fluctuations do obey linear dynamics despite the lack of a superposition 
principle.

 In the $f(R)$ case, saturation of the scalar source occurs
as the field itself saturates in value inside an overdense region, $\phi \rightarrow 0$.  More specifically, field saturation occurs when the
gravitational potential of an object exceeds the value of the background field $\bar\phi$.
 
\begin{figure}[htb]
\vspace{9pt}
\includegraphics[width=0.45\textwidth]{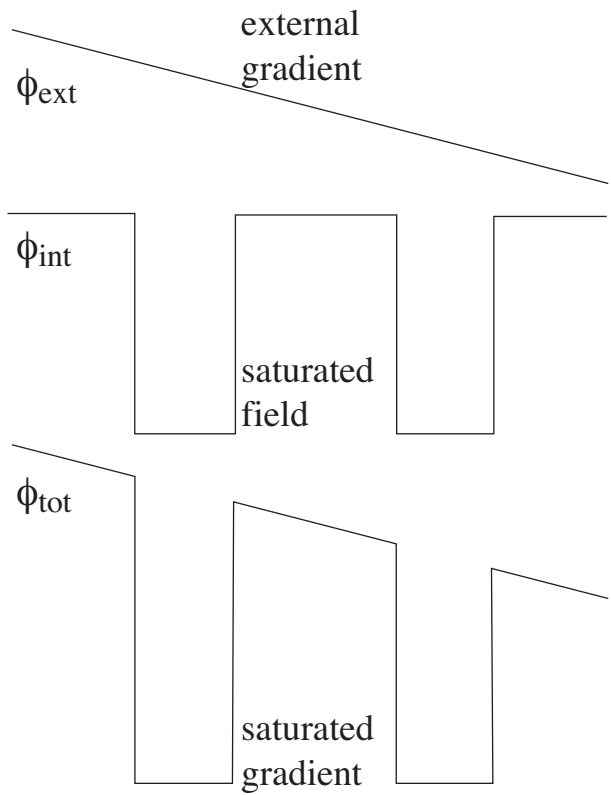}
\caption{\footnotesize Non-linear saturation of field gradients in $f(R)$ (exaggerated for effect).
For a screened object, the internal field $\phi_{\rm int}$ generated by the object does not 
superimpose with an external field $\phi_{\rm ext}$.  In particular, the field inside the object
loses knowledge of any exterior gradient and the enhanced gravitational
force it implies.  Screened objects are by definition not test bodies with respect 
to the field but still respond gravitationally to the total field $\phi_{\rm tot}$ and hence
the metric.}
\label{fig:saturated}
\end{figure}

A thought example where all of the matter resides in objects with such deep 
gravitational potentials exposes some counterintuitive properties of the lack of 
superposition.
Since the field contribution is suppressed then in all regions containing the matter, 
their contribution
to the total gravitational potential in Eq.~(\ref{eqn:modpoisson}) reflects ordinary
gravity.  Thus the large scale gravitational potential will not follow the predictions
of linear theory with modified gravity.   Likewise, the motion of these collapsed objects would also
not follow the large-scale external gradients of the field (set say by the clustering of
voids).   The saturation of the field internal to the object would eliminate the
influence of any external gradient for particles within the object (see Fig.~\ref{fig:saturated}).

This effect can be phrased as an apparent equivalence principle violation between
saturated and unsaturated objects as pointed out by \cite{Hui:2009kc}. That a saturated object
requires a non-linear field response means that
 by definition saturated objects considered as a whole are {\it not} test particles.    
 Note that the matter that
composes these objects still are test particles and so these still respond to the
total metric including the non-superimposable internal and external field contributions.
As such, there is no fundamental equivalence principle violation since matter,
regardless of its composition, will always fall on geodesics of the total, local metric.

In the cosmological $f(R)$ simulations for cases with force law deviations in the 
linear regime, the dark matter halos where most of the matter resides do not
possess deep enough gravitational potentials to be fully saturated.  In such
cases, the linear theory predictions are valid and apparent equivalence principle
violations between dark matter halos of different mass are difficult to identify.

In the DGP case, the non-linearity involves second derivatives of the field.
Let us first consider the case where the the $\nabla_i\nabla_j \phi$ and
$\nabla^2\phi$ terms are comparable, and for simplicity proportional  \cite{Khoury:2009tk},
\begin{equation}
N[\phi] = {r_c^2 \over a^4}s (\nabla^2 \phi)^2 \,.
\end{equation}
For example, the interior field of an isolated spherically symmetric
 tophat density configuration 
would have $\phi = c_1 r^2 + c_2$ and hence $s=2/3$ \cite{LueScoSta04,Khoury:2009tk}.

The non-linear field equation then becomes an algebraic relationship between
the scalar source and the density with the solution \cite{LueScoSta04} 
\begin{equation}
\nabla^2 \phi \approx {a^2 \over 2g r_c^2 s} \left[ \sqrt{1 + 32 \pi G g^2r_c^2 s \bar
\rho_m \Delta_m }-1\right]\,.
\end{equation}
At low density $\nabla^{2}\phi = g(a) a^2 (8\pi G \bar \rho_m \Delta_{m})$ as desired whereas at high density the source
is suppressed
$\nabla^{2}\phi\propto \Delta_{m}^{1/2}$.
Given the DGP form for $g$,  the threshold for the change in behavior scales as
$\Delta_m \approx 1/s$ and so the suppression of the field source is determined
by the non-linearity of the local density contrast \cite{Koyama:2007ih}. 
Contrast this with $f(R)$ where the chameleon transition is linked
to the local gravitational potential.

In the modified Poisson equation (\ref{eqn:modpoisson}), 
this looks like a density dependent Newton constant.
If this were the whole story, DGP would exhibit much stronger violations of
superposition than $f(R)$.   Since much of the total mass of the universe is in 
small collapsed objects with $\Delta_m>1$, these sources would contribute
less to the large scale modifications to gravity than their average would imply as
in our thought example above.
This is problematic for $N$-body simulations based on this approximation
\cite{Khoury:2009tk}.
Even the large scale (``linear") predictions
depend on the resolution: as more and more of the mass is resolved into collapsed
objects the modification to forces on large scales will be reduced.

Fortunately, this is not the case for the full DGP non-linear term.   Returning to the
isolated source case, the field configuration far away from the source (beyond the Vainshtein
radius) reveals an
unscreened source due to the $\nabla_i\nabla_j\phi$ terms  \cite{Khoury:2009tk,Hui:2009kc}.   Hence as long as
 the linear regime scales are larger than the Vainshtein radius
of the collapsed objects, the lack of a superposition principle does not violate
linear theory predictions.  
This fact is also related to the absence of an
apparent equivalence principle violation in DGP \cite{Hui:2009kc}.  

Finally,
 it is interesting
to note that for a planar configuration $N[\phi]=0$ for DGP and there is no non-linear
suppression even when $\Delta_m \gg 1$.   
This has the interesting consequence that cosmological structure that is planar
or sheetlike exhibits the largest modification to gravity.  
These counterterms that suppress the non-linearity for planar configurations
are a generic feature of models with ``galileon" shift symmetry \cite{Nicolis:2008in}.

\section{DISCUSSION}

Perhaps the primary lesson of the worked examples is that gravity is fragile and preserving
what we know about gravity already poses severe restrictions on a viable modification that
explains cosmic acceleration.   On the other hand, this fragility provides form 
to parametric approaches and insight on how to design cosmological tests that best complement our knowledge
of gravity.  

 Demanding a metric theory where the matter obeys a conservation law
 requires that deviations take the form of a dark energy component
under ordinary gravity.   The most important parameter for modified gravity 
in this language is
an effective dark energy anisotropic stress that is coupled directly to the lensing potential.
The resulting
 phenomenology is therefore distinct from typical physical dark energy models.
It is important to note that the large scale Friedmann and small scale Newtonian dynamics  implied by the modifications
differ.  They do so for the same reason that dark energy cannot be considered smooth 
relative to the dark matter on horizon scales and above: conservation of energy-momentum.  The parameterization 
given in \cite{HuSaw07b} spans both regimes and allows linear theory tools for
general relativity to be used for modified gravity \cite{Hu08}.

There does not yet exist a complete parameterization of the non-linear phenomenology of modified
gravity but any viable description must contain a mechanism to suppress the modification
locally.   Density based approaches work well for the DGP model but less well for
the potential driven $f(R)$ model.  
More importantly, the DGP and $f(R)$ models provide specific examples
of this suppression.  In both cases, the modification is 
mediated by a scalar that obeys a non-linear field equation coupled to density fluctuations
that can be solved numerically
via relaxation techniques \cite{Oyaizu:2008sr}.    Non-linearities in the
field equation prevent superposition of field solutions given the matter.  In particular
they allow saturation effects where the modifications are suppressed within the
high overdensities where local tests of gravity are performed.   

The drawback of
the lack of superposition is that  for cosmological tests, the field equations must
be solved jointly with the matter in $N$-body simulations.  In principle this caveat applies
even to large scales in the supposedly linear regime.  The DGP and $f(R)$ models however
provide examples where the
deviations from linear theory predictions there are small.

Likewise, the motion of matter within bound  objects, while still responding gravitationally only to the metric,
cannot always be thought of as a superposition of external and internal influences.  
These features caution against the use of simple scalings that take the linear
behavior of modified gravity and seek to predict non-linear cosmological observables.

\bibliographystyle{elsart-num}
\bibliography{lessons}

\def\eprinttmppp@#1arXiv:@{#1}
\providecommand{\arxivlink[1]}{\href{http://arxiv.org/abs/#1}{arXiv:#1}}
\def\eprinttmp@#1arXiv:#2 [#3]#4@{\ifthenelse{\equal{#3}{x}}{\ifthenelse{
\equal{#1}{}}{\arxivlink{\eprinttmppp@#2@}}{\arxivlink{#1}}}{\arxivlink{#2}
  [#3]}}
\providecommand{\eprintlink}[1]{\eprinttmp@#1arXiv: [x]@}
\providecommand{\eprint}[1]{\eprintlink{#1}}
\providecommand{\eprintmod}[1][XXXX.XXXX]{\eprintlink{#1}}
\providecommand{\adsurl}[1]{\href{#1}{ADS}}
\newcommand{\apj}{Astrophys. J.}
\newcommand{\prd}{Phys. Rev. D}
\begin{thebibliography}{10}
\expandafter\ifx\csname url\endcsname\relax
  \def\url#1{\texttt{#1}}\fi
\expandafter\ifx\csname urlprefix\endcsname\relax\def\urlprefix{URL }\fi

\bibitem{Albrecht:2006um}
A.~J. Albrecht, et~al. \eprintmod[arXiv:astro-ph/0609591].

\bibitem{Albrecht:2009ct}
A.~J. Albrecht, et~al. \eprintmod[arXiv:0901.0721].

\bibitem{Frieman:2008sn}
J.~Frieman, M.~Turner, D.~Huterer, Ann. Rev. Astron. Astrophys. 46 (2008)
  385--432,  \eprintmod[arXiv:0803.0982].

\bibitem{Linder:2008pp}
E.~V. Linder, Rept. Prog. Phys. 71 (2008) 056901,  \eprintmod[arXiv:0801.2968].

\bibitem{Caretal03}
S.~M. Carroll, V.~Duvvuri, M.~Trodden, M.~S. Turner, Phys. Rev. D70 (2004)
  043528,  \eprintmod[arXiv:astro-ph/0306438].

\bibitem{NojOdi03}
S.~Nojiri, S.~D. Odintsov, Phys. Rev. D68 (2003) 123512,
  \eprintmod[arXiv:hep-th/0307288].

\bibitem{Capozziello:2003tk}
S.~Capozziello, S.~Carloni, A.~Troisi, Recent Res. Dev. Astron. Astrophys. 1
  (2003) 625,  \eprintmod[arXiv:astro-ph/0303041].

\bibitem{DvaGabPor00}
G.~R. Dvali, G.~Gabadadze, M.~Porrati, Phys. Lett. B485 (2000) 208--214,
  \eprintmod[arXiv:hep-th/0005016].

\bibitem{Kunz:2006ca}
M.~Kunz, D.~Sapone, Phys. Rev. Lett. 98 (2007) 121301,
  \eprintmod[arXiv:astro-ph/0612452].

\bibitem{HuSaw07b}
W.~Hu, I.~Sawicki, Phys. Rev. D76 (2007) 104043,  \eprintmod[arXiv:0708.1190].

\bibitem{Hu98}
W.~Hu, Astrophys. J. 506 (1998) 485--494,  \eprintmod[arXiv:astro-ph/9801234].

\bibitem{Linder:2007hg}
E.~V. Linder, R.~N. Cahn, Astropart. Phys. 28 (2007) 481,
  \eprintmod[arXiv:astro-ph/0701317].

\bibitem{HutLin07}
D.~Huterer, E.~V. Linder, Phys. Rev. D75 (2007) 023519,
  \eprintmod[arXiv:astro-ph/0608681].

\bibitem{IshUpaSpe06}
M.~Ishak, A.~Upadhye, D.~N. Spergel, Phys. Rev. D74 (2006) 043513,
  \eprintmod[arXiv:astro-ph/0507184].

\bibitem{WanHuiMayHai07}
S.~Wang, L.~Hui, M.~May, Z.~Haiman, Phys. Rev. D76 (2007) 063503,
  \eprintmod[arXiv:0705.0165].

\bibitem{Mortonson:2008qy}
M.~J. Mortonson, W.~Hu, D.~Huterer, Phys. Rev. D79 (2009) 023004,
  \eprintmod[arXiv:0810.1744].

\bibitem{Alam:2008at}
U.~Alam, V.~Sahni, A.~A. Starobinsky \eprintmod[arXiv:0812.2846].

\bibitem{Hearin:2009hz}
A.~P. Hearin, A.~R. Zentner \eprintmod[arXiv:0904.3334].

\bibitem{SonHuSaw06}
Y.~{Song}, W.~{Hu}, I.~{Sawicki}, \prd 75 (2007) 044004,
  \eprintmod[arXiv:astro-ph/0610532].

\bibitem{SawSonHu06}
I.~{Sawicki}, Y.~{Song}, W.~{Hu}, \prd 75 (2006) 064002,
  \eprintmod[arXiv:astro-ph/0606285].

\bibitem{cardoso:08}
A.~Cardoso, K.~Koyama, S.~S. Seahra, F.~P. Silva, Phys. Rev. D77 (2008) 083512,
   \eprintmod[arXiv:0711.2563].

\bibitem{Amendola:2007rr}
L.~Amendola, M.~Kunz, D.~Sapone, JCAP 0804 (2008) 013,
  \eprintmod[arXiv:0704.2421].

\bibitem{Uzan:2006mf}
J.-P. Uzan, Gen. Rel. Grav. 39 (2007) 307--342,
  \eprintmod[arXiv:astro-ph/0605313].

\bibitem{CalCooMel07}
R.~Caldwell, A.~Cooray, A.~Melchiorri, Phys. Rev. D76 (2007) 023507,
  \eprintmod[arXiv:astro-ph/0703375].

\bibitem{HuEis99}
W.~{Hu}, D.~J. {Eisenstein}, \prd 59 (1999) 083509,
  \eprintmod[arXiv:astro-ph/9809368].

\bibitem{Ber06}
E.~Bertschinger, Astrophys. J. 648 (2006) 797--806,
  \eprintmod[arXiv:astro-ph/0604485].

\bibitem{Sawicki:2006jj}
I.~Sawicki, Y.-S. Song, W.~Hu, Phys. Rev. D75 (2007) 064002,
  \eprintmod[arXiv:astro-ph/0606285].

\bibitem{SonSawHu06}
Y.~{Song}, I.~{Sawicki}, W.~{Hu}, \prd 75 (2006) 064003,
  \eprintmod[arXiv:astro-ph/0606286].

\bibitem{Fang:2008kc}
W.~Fang, et~al., Phys. Rev. D78 (2008) 103509,  \eprintmod[arXiv:0808.2208].

\bibitem{Daniel:2008et}
S.~F. Daniel, R.~R. Caldwell, A.~Cooray, A.~Melchiorri, Phys. Rev. D77 (2008)
  103513,  \eprintmod[arXiv:0802.1068].

\bibitem{Jain:2007yk}
B.~Jain, P.~Zhang, Phys. Rev. D78 (2008) 063503,  \eprintmod[arXiv:0709.2375].

\bibitem{AmiWagBla07}
M.~A. Amin, R.~V. Wagoner, R.~D. Blandford \eprintmod[arXiv:0708.1793].

\bibitem{Hu08}
W.~Hu, Phys. Rev. D77 (2008) 103524,  \eprintmod[arXiv:0801.2433].

\bibitem{Fang:2008sn}
W.~Fang, W.~Hu, A.~Lewis, Phys. Rev. D78 (2008) 087303,
  \eprintmod[arXiv:0808.3125].

\bibitem{Lombriser:2009xg}
L.~Lombriser, W.~Hu, W.~Fang, U.~Seljak \eprintmod[arXiv:0905.1112].

\bibitem{Afshordi:2008rd}
N.~Afshordi, G.~Geshnizjani, J.~Khoury \eprintmod[arXiv:0812.2244].

\bibitem{Koyama:2009me}
K.~Koyama, A.~Taruya, T.~Hiramatsu \eprintmod[arXiv:0902.0618].

\bibitem{Rapetti:2008rm}
D.~Rapetti, S.~W. Allen, A.~Mantz, H.~Ebeling \eprintmod[arXiv:0812.2259].

\bibitem{KhoWel04}
J.~Khoury, A.~Weltman, Phys. Rev. D69 (2004) 044026,
  \eprintmod[arXiv:astro-ph/0309411].

\bibitem{Oyaizu:2008sr}
H.~Oyaizu, Phys. Rev. D78 (2008) 123523,  \eprintmod[arXiv:0807.2449].

\bibitem{Oyaizu:2008tb}
H.~Oyaizu, M.~Lima, W.~Hu, Phys. Rev. D78 (2008) 123524,
  \eprintmod[arXiv:0807.2462].

\bibitem{Schmidt:2009sg}
F.~Schmidt \eprintmod[arXiv:0905.0858].

\bibitem{Schmidt:2008tn}
F.~Schmidt, M.~V. Lima, H.~Oyaizu, W.~Hu \eprintmod[arXiv:0812.0545].

\bibitem{Hui:2009kc}
L.~Hui, A.~Nicolis, C.~Stubbs \eprintmod[arXiv:0905.2966].

\bibitem{Khoury:2009tk}
J.~Khoury, M.~Wyman \eprintmod[arXiv:0903.1292].

\bibitem{LueScoSta04}
A.~Lue, R.~Scoccimarro, G.~D. Starkman, Phys. Rev. D69 (2004) 124015,
  \eprintmod[arXiv:astro-ph/0401515].

\bibitem{Koyama:2007ih}
K.~Koyama, F.~P. Silva, Phys. Rev. D75 (2007) 084040,
  \eprintmod[arXiv:hep-th/0702169].

\bibitem{Nicolis:2008in}
A.~Nicolis, R.~Rattazzi, E.~Trincherini \eprintmod[arXiv:0811.2197].

\end{thebibliography}

\end{document}